\documentclass{appolb}
\usepackage{graphicx}
\usepackage{mciteplus}
\usepackage{epstopdf}
\usepackage{ulem}
\usepackage{hyperref}

\begin{document}
\title{NuWro Monte Carlo generator of neutrino interactions -- First Electron Scattering Results%
\thanks{Presented at the XXXIX International Conference of Theoretical Physics ``Matter to the Deepest'' by J. \.{Z}muda, jakub.zmuda@ift.uni.wroc.pl}%
}
\author{ Jakub \.{Z}muda, Krzysztof M. Graczyk, Cezary Juszczak, Jan T. Sobczyk
\address{Institute of Theoretical Physics, University of Wroclaw, Poland}
}
\maketitle
\begin{abstract}
NuWro Monte Carlo generator of events is presented. It is a numerical environment containing all necessary ingredients to simulate interactions of neutrinos with nucleons and nuclei in realistic experimental situation in wide neutrino energy range.  It can be used both for data analysis as well
as studies of nuclear effects in neutrino interactions. The first results and functionalities of eWro - module of NuWro dedicated  to  electron nucleus scattering - are also presented.  
\end{abstract}
\PACS{25.30.Pt, 13.15.+g}
  
\section{Introduction}

For the past twenty years there has been a growing interest in the neutrino oscillations. Since the confirmation
of this phenomenon in Super-Kamiokande and Sudbury experiments a lot of effort has been made towards precise
measurement of lepton mixing angles and mass differences. More challenging studies of neutrino mass hierarchy,
leptonic CP violation as well as existence of sterile neutrino are addressed by a series of new experiments. We pay here a special
attention to the ones using accelerator neutrino beam sources, including MINOS+, T2K, NovA, MicroBooNE and planned DUNE and Hyper-Kamiokande. A large part of systematic uncertainties
comes from a lack of precise knowledge about neutrino-nucleus interaction physics.
Complexity of this problem is largely due to difficulties in modeling nuclear structure effects in large energy range spanned by neutrino beams. 
This requires the use of different theoretical formulations for various pieces of the phase space,
starting from the nonrelativistic quantum mechanics through effective hadronic field theories up to quark jet fragmentation routines
for deep inelastic scattering. All created and participating particles propagate through
atomic nucleus, where they are subject to strong final state interactions (FSI). It is desirable that the systematic error in new oscillation experiments be reduced to 1-3\%. It is a very challenging goal because the knowledge of neutrino-nucleus cross sections is not better than 10-20\%. Beside the oscillation studies accelerator neutrinos are a probe to test weak interaction of hadrons and atomic nuclei, which makes the
physical program even  richer and more interesting.
\begin{figure}[h]
\centerline{%
\includegraphics[width=1.05\textwidth,angle=-90]{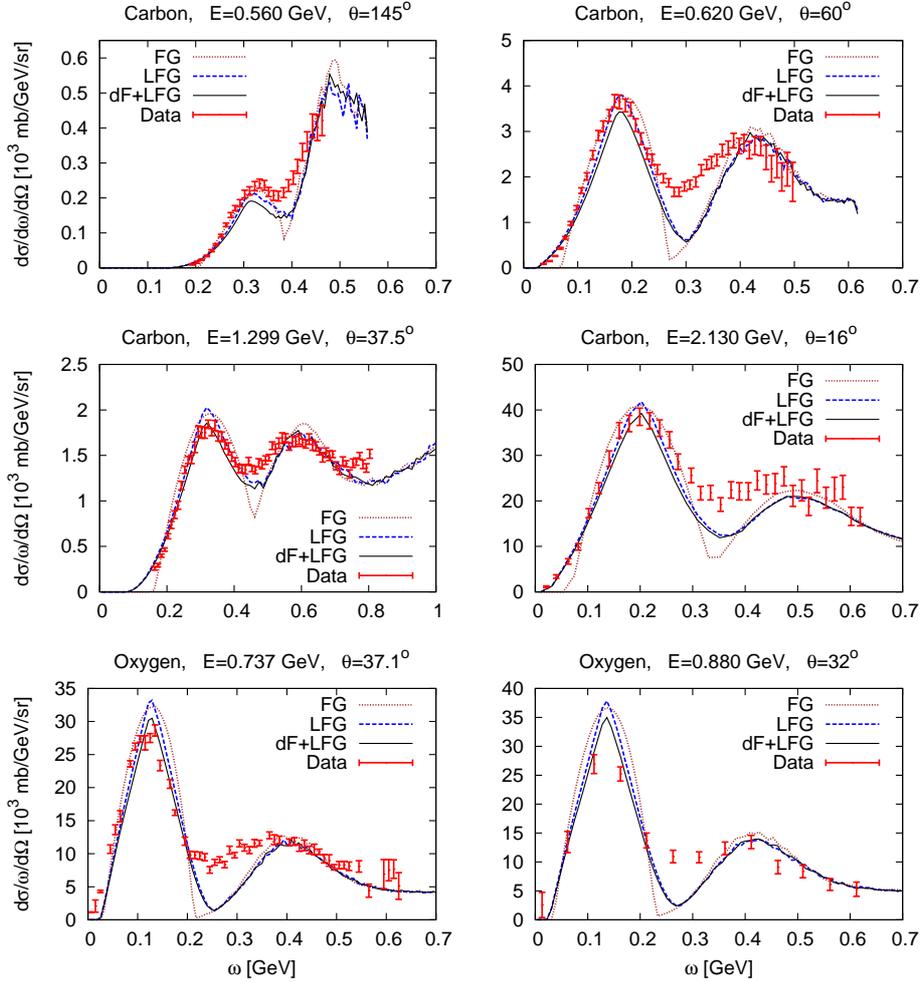}}
\caption{Differential cross sections  for electron scattering off carbon and oxygen obtained within eWro (for various beam energies, $E$, and scattering angles, $\theta$).  The curves correspond to FG, LFG and LFG with de Forest treatment of binding energy (dF+LFG). 
The data are taken from \cite{Barreau:1983htshort} (top), \cite{Sealock:1989nxshort} (middle left),
\cite{Bagdasaryan:1988hpshort} (middle right), \cite{O'Connell:1987agshort} (bottom left), \cite{Anghinolfi:1996vm} (bottom right).}
\label{Fig:F2H}
\end{figure}


Monte-Carlo generator of events such as NuWro \cite{Golan:2012wx}
 helps in analysis of experimental measurements. 
NuWro has been developed since $\sim 2004$. 
Even though, it is not an official Monte Carlo of any experiment, it is widely used as a tool for development and testing of new physical models in MC generators. This includes {\it e.g.}\ implementation of the IFIC model of two-nucleon currents
\cite{Nieves:2011pp} and the Berger-Sehgal model of coherent pion production \cite{Berger:2008xs}. NuWro is also used as a benchmark and reference for experimental collaborations \cite{Acciarri:2014isz,Aliaga:2015wva}.
Overview and references to other MC generators (such as GENIE, NEUT and GiBUU) can be found in  \cite{Agashe:2014kda}.

Recently a new electron scattering simulation module has been added to NuWro. It allows for comparisons with accurate
electron-nucleus scattering data and extensive tests of implemented physical models. This makes NuWro a fully-fledged and versatile
tool for theoretical and experimental physicists. Its main interaction channels and implemented physical models
will be outlined in following sections.

\section{NuWro}
NuWro generator is capable of simulating neutrino interactions taking into account beam profile and composition, detailed detector geometry 
as well as FSI in the nuclear target. 
It can be applied to a  neutrino energies ranging from around  100 MeV, implied by the requirement of validity of the impulse approximation, up to the TeV energy range.
In between the possible physical channels change from charged-current quasielastic (CCQE), neutral current elastic (NCE) scattering, through multi-nucleon meson exchange currents (MEC),
coherent (COH) and resonant (RES) pion production up to deep inelastic scattering (DIS), which stands here for processes more inelastic
from RES. A description of nuclear system in NuWro can be chosen from a variety of options: global or local Fermi gas models (FG or LFG), hole spectral function \cite{Benhar:1994hw,Rohe:2004dz,Rohe:2006er}, effective momentum and density dependent potential \cite{Juszczak:2005wk}. The object-oriented
C++ code is very flexible and allows one to easily introduce new theoretical models and modify the existing ones. The choice of physical
interaction channels, 
 model parameters and variants, such as nucleon and resonance form factor sets,
is made within the universal parameter file. The output is written in three separate files containing respectively: the simulations parameter set,
the  integrated cross sections for each interaction mode and target, and the ROOT tree of equally weighted events with
information about initial, intermediate and final state particles. 
The NuWro code is open source and can be downloaded from \cite{NuWro}. A prototype {\sl NuWro Online} service is also available there.

The CCQE and NCE scattering processes are very important for neutrino experiments with beam energy range up to few GeV \cite{Abe:2011ks, AguilarArevalo:2007it}.
In NuWro the FG/LFG model has been improved by implementing relativistic ring random phase approximation (RPA) \cite{Graczyk:2003ru} with effective nucleon mass (only for CCQE).  
NuWro contains also an implementation of the spectral function approach  \cite{Ankowski:2007uy}.  For both CCQE and NCE interactions one can choose between multiple form factor sets describing  neutrino-nucleon vertex structure.

Importance of MEC in neutrino experiments has been first pointed out by Martini {\textit{et al.}} in \cite{Martini:2009uj} in connection with MiniBooNE
CCQE cross section measurement publication \cite{AguilarArevalo:2010zc}. The latter analysis did not consider the MEC contribution, which lead to abnormally large measured
nucleon axial mass parameter, whereas the models including MEC tend to reproduce its standard value from the MiniBooNE data \cite{Nieves:2011yp}. In NuWro
there are two MEC implementations available: the effective transverse enhancement model \cite{Bodek:2011ps}, which parametrizes MEC effects as an
enlargement of magnetic nucleon form factors, and microscopic IFIC model \cite{Nieves:2011pp} (only for charge current interactions).

The RES channel includes pion production processes in the invariant mass range $W<1.6$ GeV. It consists of the $\Delta(1232)$ resonance excitation \cite{Graczyk:2009qm}
and effective background extrapolated from the DIS contribution down to $W=1.4$ GeV. There are multiple $\Delta$ form factor sets available including the fits to ANL/BNL bubble chamber data from \cite{Graczyk:2009qm}.
The lack of heavier resonances is justified by the quark-hadron duality hypothesis \cite{Graczyk:2005uv} and Fermi motion effects, which wash out the distinct resonance signal.
Recently pion angular correlations measured in the ANL and BNL experiments have been included \cite{Radecky:1981fnshort,Kitagaki:1986ctshort} 
as well as an approximate implementation of the $\Delta(1232)$ self-energy in nuclear matter from \cite{Oset:1987re}.

Coherent pion production process leaves the nucleus in its ground state. In NuWro it is simulated using two phenomenological PCAC based models described in Rein-Sehgal \cite{Rein:1982pf} and Berger-Sehgal \cite{Berger:2008xs} papers, the latter seemingly closer to recent MINERvA data \cite{Higuera:2014azj}.

The DIS channel contains inelastic interactions in a region of $W>1.6$~GeV. The overall cross section is evaluated using the Bodek-Yang corrections \cite{Bodek:2002vp}. Hadronic final states are obtained using PYTHIA 6 hadronization routines \cite{Jarek}. In a region of $W\in (1.4, 1.6)$~GeV a smooth transition from RES to DIS is done.  

The FSI algorithm in NuWro is based on  intranuclear cascade \cite{Metropolis:1958sb}. It uses phenomenological and experimental cross sections for elastic and inelastic pion-nucleon
and nucleon-nucleon interactions, including the Oset model from  \cite{Salcedo:1987md} and finite $\Delta$ life-time effects. Nucleon-nucleon effective cross section includes in-medium
modifications following the approach of Pandharipande-Pieper \cite{Pandharipande:1992zz}. More detailed overview of NuWro FSI effects can be found in \cite{Tomek}.

\section{Electron scattering in NuWro}

The work is under way on eWro - a new module for generating the electron-nucleus interactions. 
Due to the abundance and good precision of electron scattering data eWro can serve as a testing ground of nuclear physics models which are common in eWro and NuWro.

Currently in eWro nucleus is modeled within FG/LFG.  The quasielastic and single pion production  processes are implemented. The latter channel  contains $\Delta(1232)$ resonance and nonresonant background from \cite{Hernandez:2007qq,Graczyk:2014dpa}
and implementation of nonperturbative $\Delta(1232)$ self-energy following  \cite{Oset:1987re}. The RES model of eWro differs  from that of NuWro with respect of treatment of non-resonant background.

In Fig.  \ref{Fig:F2H} eWro predictions for inelastic electron scattering off carbon and oxygen targets are  presented. We show results for FG and LFG models of the ground state of the nucleus. The binding energy following the de Forest prescription \cite{DeForest:1983vc} is taken into account. 
In most of the plots the quasielastic and $\Delta (1232)$ peaks are clearly visible. To describe $N\to \Delta(1232)$ transition  recent  fits of electromagnetic form
factors from  \cite{cetup14} has been used. The $\Delta$ propagator is fully dressed in medium effects following the approach described in \cite{Oset:1987re}.

\section{Summary}

NuWro is a versatile tool to study lepton-nucleus interactions and produce MC data samples for experimental analysis purposes. It covers most important
interaction modes. It is a flexible code ready for further developments and use in actual data analysis.\\
This work was supported by NCN Grant No. UMO-2011/M/ST2/02578.

\bibliographystyle{polonica}
\bibliography{bibdrat}

\begin{thebibliography}{10}

\bibitem{Barreau:1983htshort}
P.~Barreau, et~al.,
\newblock {\em Nucl. Phys. A}
\newblock {\bf  402}, 515 (1983).

\bibitem{Sealock:1989nxshort}
R.M. Sealock, et~al.,
\newblock {\em Phys. Rev. Lett.}
\newblock {\bf  62}, 1350 (1989).

\bibitem{Bagdasaryan:1988hpshort}
D.~S. Bagdasaryan, et~al.,
\newblock YERPHI-1077-40-88 1988.

\bibitem{O'Connell:1987agshort}
J.S. O'Connell, et~al.,
\newblock {\em Phys. Rev. C}
\newblock {\bf  35}, 1063 (1987).

\bibitem{Anghinolfi:1996vm}
M.~Anghinolfi, et~al.,
\newblock {\em Nucl. Phys.}
\newblock {\bf  A602}, 405 (1996).

\bibitem{Golan:2012wx}
T.~Golan, C.~Juszczak, J.~T. Sobczyk,
\newblock {\em Phys.\ Rev.\ C}
\newblock {\bf  86}, 015505 (2012).

\bibitem{Nieves:2011pp}
J.~Nieves, I.~Ruiz~Simo, M.~J. Vicente~Vacas,
\newblock {\em ibid.}
\newblock {\bf  83}, 045501 (2011).

\bibitem{Berger:2008xs}
Ch. Berger, L.M. Sehgal,
\newblock {\em Phys.Rev.}
\newblock {\bf  D79}, 053003 (2009).

\bibitem{Acciarri:2014isz}
R.~Acciarri, et~al.,
\newblock {\em Phys. Rev.}
\newblock {\bf  D89}, 112003 (2014).

\bibitem{Aliaga:2015wva}
T.~Le, et~al.,
\newblock {\em Phys. Lett.}
\newblock {\bf  B749}, 130 (2015).

\bibitem{Agashe:2014kda}
K.~A. Olive, et~al.,
\newblock {\em Chin. Phys.}
\newblock {\bf  C38}, 090001 (2014).

\bibitem{Benhar:1994hw}
O.~Benhar, A.~Fabrocini, S.~Fantoni, I.~Sick,
\newblock {\em Nucl.\ Phys.\ A}
\newblock {\bf  579}, 493 (1994).

\bibitem{Rohe:2004dz}
D.~Rohe, C.~S. Armstrong, R.~Asaturyan, O.~K. Baker, S.~Bueltmann, C.~Carasco,
  D.~Day, R.~Ent~{\it et al.},
\newblock {\em Phys.\ Rev.\ Lett.}
\newblock {\bf  93}, 182501 (2004).

\bibitem{Rohe:2006er}
D.~Rohe~{\it et al.},
\newblock {\em Nucl.\ Phys.\ Proc.\ Suppl.}
\newblock {\bf  159}, 152 (2006).

\bibitem{Juszczak:2005wk}
C.~Juszczak, J.~A. Nowak, J.~T. Sobczyk,
\newblock {\em Eur. Phys. J.}
\newblock {\bf  C39}, 195 (2005).

\bibitem{NuWro}
{NuWro repository}
\newblock \url{http://borg.ift.uni.wroc.pl/nuwro/}.

\bibitem{Abe:2011ks}
K.~Abe, et~al.,
\newblock {\em Nucl. Instrum. Meth.}
\newblock {\bf  A659}, 106 (2011).

\bibitem{AguilarArevalo:2007it}
A.A. Aguilar-Arevalo, et~al.,
\newblock {\em Phys.Rev.Lett.}
\newblock {\bf  98}, 231801 (2007).

\bibitem{Graczyk:2003ru}
K.~M. Graczyk, J.~T. Sobczyk,
\newblock {\em Eur.\ Phys.\ J.\ C}
\newblock {\bf  31}, 177 (2003).

\bibitem{Ankowski:2007uy}
A.~M. Ankowski, J.~T. Sobczyk,
\newblock {\em Phys.\ Rev.\ C}
\newblock {\bf  77}, 044311 (2008).

\bibitem{Martini:2009uj}
M.~Martini, M.~Ericson, G.~Chanfray, J.~Marteau,
\newblock {\em ibid.}
\newblock {\bf  {\bf 80}}, 065501 (2009).

\bibitem{AguilarArevalo:2010zc}
A.~A. Aguilar-Arevalo, et~al.,
\newblock {\em Phys.\ Rev.\ D}
\newblock {\bf  {\bf 81}}, 092005 (2010).

\bibitem{Nieves:2011yp}
J.~Nieves, I.~Ruiz~Simo, M.~J. Vicente~Vacas,
\newblock {\em Phys.\ Lett.\ B}
\newblock {\bf  {\bf 707}}, 72 (2012).

\bibitem{Bodek:2011ps}
H.~S. Bodek, A.~Budd, M.~E. Christy,
\newblock {\em Eur.\ Phys.\ J.\ C}
\newblock {\bf  71}, 1726 (2011).

\bibitem{Graczyk:2009qm}
K.~M. Graczyk, D.~Kielczewska, P.~Przewlocki, J.~T. Sobczyk,
\newblock {\em Phys.\ Rev.\ D}
\newblock {\bf  80}, 093001 (2009).

\bibitem{Graczyk:2005uv}
K.~M. Graczyk, C.~Juszczak, J.~T. Sobczyk,
\newblock {\em Nucl.\ Phys.\ A}
\newblock {\bf  781}, 227 (2007).

\bibitem{Radecky:1981fnshort}
G.~M. Radecky, et~al.,
\newblock {\em Phys.\ Rev.\ D}
\newblock {\bf  {\bf 26}}, 3297 (1982), {\em [Erratum-ibid.\ D {\bf 26} (1982)
  3297]}.

\bibitem{Kitagaki:1986ctshort}
T.~Kitagaki, et~al.,
\newblock {\em Phys. Rev. D}
\newblock {\bf  34}, 2554 (1986).

\bibitem{Oset:1987re}
E.~Oset, L.~L. Salcedo,
\newblock {\em Nucl.\ Phys.\ A}
\newblock {\bf  {\bf 468}}, 631 (1987).

\bibitem{Rein:1982pf}
D.~Rein, L.~M. Sehgal,
\newblock {\em Nucl.\ Phys.\ B}
\newblock {\bf  223}, 29 (1983).

\bibitem{Higuera:2014azj}
A.~Higuera, et~al.,
\newblock {\em Phys.Rev.Lett.}
\newblock {\bf  113}, 261802 (2014).

\bibitem{Bodek:2002vp}
A.~Bodek, U.~K. Yang,
\newblock {\em Nucl.\ Phys.\ Proc.\ Suppl.}
\newblock {\bf  112}, 70 (2002).

\bibitem{Jarek}
J.~Nowak,
\newblock {\em {Construction of neutrino event generator}},
\newblock PhD thesis, {Uniwersity of Wroclaw} 2006,
\newblock {\em {in Polish}}.

\bibitem{Metropolis:1958sb}
N.~Metropolis, R.~Bivins, M.~Storm, J.~M. Miller, G.~Friedlander, Anthony
  Turkevich,
\newblock {\em Phys. Rev.}
\newblock {\bf  110}, 204 (1958).

\bibitem{Salcedo:1987md}
L.~L. Salcedo, E.~Oset, M.~J. Vicente-Vacas, C.~Garcia-Recio,
\newblock {\em Nucl.\ Phys.\ A}
\newblock {\bf  484}, 557 (1988).

\bibitem{Pandharipande:1992zz}
V.~R. Pandharipande, Steven~C. Pieper,
\newblock {\em Phys. Rev.}
\newblock {\bf  C45}, 791 (1992).

\bibitem{Tomek}
T.~Golan,
\newblock {\em {Modeling nuclear effects in NuWro Monte Carlo neutrino event
  generator}},
\newblock PhD thesis, {University of Wroclaw} 2014.

\bibitem{Hernandez:2007qq}
E.~Hernandez, J.~Nieves, M.~Valverde,
\newblock {\em Phys.\ Rev.\ D}
\newblock {\bf  {\bf 76}}, 033005 (2007).

\bibitem{Graczyk:2014dpa}
K.~M. Graczyk, J.~\.{Z}muda, J.T. Sobczyk,
\newblock {\em Phys.Rev.}
\newblock {\bf  D90}, 093001 (2014).

\bibitem{DeForest:1983vc}
T.~De~Forest,
\newblock {\em Nucl.\ Phys.\ A}
\newblock {\bf  392}, 232 (1983).

\bibitem{cetup14}
J.~\.Zmuda, K.~M. Graczyk,
\newblock arXiv:1501.03086v4 2015.

\end{thebibliography}

\end{document}